\documentclass[%
reprint,
superscriptaddress,
amsmath,amssymb,
prb,
]{revtex4-1}

\usepackage{graphicx}% Include figure files
\usepackage{subfigure}
\usepackage{dcolumn}% Align table columns on decimal point
\usepackage{bm}% bold math
\usepackage{bbold}
\usepackage{feynmp}									% For Feynman diagram scripts
\usepackage{color}

\begin{document}
\setlength{\unitlength}{1mm}		% set the scale for Feynman diagrams
\begin{fmffile}{fgraphs}				% compile Feynman-diagram declarations to 'fgraphs.mp'

%\preprint{APS/123-QED}

\title{Chemical-potential flow equations for graphene with Coulomb interactions}

\author{Christian Fr\"a\ss dorf}
\affiliation{%
Dahlem Center for Complex Quantum Systems and, Institut f\"ur Theoretische Physik, \\Freie Universit\"at Berlin, Arnimallee 14, 14195 Berlin, Germany}%

\author{Johannes E. M. Mosig}
\affiliation{%
Department of Mathematics and Statistics, University of Otago, PO Box 56, Dunedin 9054, New Zealand}%

\date{\today}%

\begin{abstract}
We calculate the chemical potential dependence of the renormalized Fermi velocity and static dielectric function for Dirac quasiparticles in graphene nonperturbatively at finite temperature. By reinterpreting the chemical potential as a flow parameter in the spirit of the functional renormalization group (fRG) we obtain a set of flow equations, which describe the change of these functions upon varying the chemical potential. In contrast to the fRG the initial condition of the flow is nontrivial and has to be calculated separately. Our results confirm that the charge carrier density dependence of the Fermi velocity is negligible, validating the comparison of the fRG calculation at zero density of Bauer \textit{et al.}, Phys.\ Rev.\ B {\bf 92}, 121409 (2015) with the experiment of Elias \textit{et al.}, Nat.\ Phys.\ {\bf 7}, 701 (2011).
\end{abstract}

\maketitle

The spectrum of free electrons in graphene is characterized by two Dirac points around which the energy disperses linearly as a function of momentum.~\cite{Wallace1947, Semenoff1984, CastroNetoGuineaNovoselovGeim2009} One important peculiarity of the linear band structure is that it leads to a vanishing density of states at these nodal points. The vanishing charge carrier density implies the absence of screening, leading to strongly enhanced corrections of the system's single-particle properties by the long-range tail of the Coulomb interaction. One-loop calculations have shown that the Fermi velocity acquires logarithmic corrections upon approaching the nodal points.~\cite{GonzalezGuineaVozmediano1993, Gonzalez1999, Vozmediano2011, KotovEtal2012} These corrections diverge precisely at the nodal points at zero temperature, which corresponds to a strongly increasing Fermi velocity.

This effect becomes most pronounced in the strong coupling regime, which is experimentally realized by freestanding graphene, where there is no screening dielectric surrounding the graphene sheet. Such an experiment has been performed recently by Elias \textit{et al.}\cite{EliasEtal2011}, verifying the theoretical prediction. Since perturbative calculations are not reliable in such a situation - the dimensionless interaction strength in freestanding graphene is about 2.2 - nonperturbative methods have been employed to address this issue theoretically. Bauer \textit{et al.}~\cite{BauerKopietz2015} used the functional renormalization group (fRG) formalism to access the strong coupling regime,~\cite{Wetterich2001, BergesTetradisWetterich2002, MetznerHonercampEtal2012, KopietzBook} finding excellent agreement with the experiment of Elias \textit{et al.} Upon closer inspection, however, the calculation of Ref.~[\onlinecite{BauerKopietz2015}] addresses a slightly different quantity than what is measured in the experiment of Ref.~[\onlinecite{EliasEtal2011}]. The theoretical calculation has been performed at zero density and equates the momentum dependent quasiparticle velocity $v(k)$ with the Fermi velocity in a system with finite carrier density at Fermi momentum $k = k_F$. The experiment, in contrast, observed the logarithmic increase of the Fermi velocity as a function of the charge carrier density. Strictly speaking these two velocities are different aspects of a more general velocity function, which depends on momentum, chemical potential and temperature. Equating the velocities of Refs.~[\onlinecite{EliasEtal2011}] and [\onlinecite{BauerKopietz2015}] requires that the carrier density dependence of the full velocity function is negligible. This identification allows one to map the momentum dependence to a density dependence, which could then be compared to the experiment. It is the goal of this paper to revisit this issue and calculate the density dependence of the renormalized momentum dependent quasiparticle velocity in order to verify this key assumption.

Standard application of the fRG to calculate the renormalization of the Fermi velocity at finite density requires a repeated solution of the truncated vertex flow equations for every value of the chemical potential. Furthermore, at finite density a renormalization of the Fermi surface under the RG flow~\cite{MetznerHonercampEtal2012} has to be accounted for.
%
%depending on the regularization scheme it can be difficult to keep track of the renormalization of the Fermi surface under the RG flow.~\cite{MetznerHonercampEtal2012} 
%
To circumvent both complications, we here use a variant of the fRG, where the chemical potential $\mu$ is interpreted as a flow parameter.~\cite{BergesEtal2003} In contrast to conventional fRG where one is only interested in the one-particle irreducible vertex functions at the end of the flow, the chemical-potential flow bears physical information for all values of the flow parameter. The solution to the chemical potential flow equations directly gives access to the full $\mu$ dependence of the vertex functions and there are no issues regarding the renormalization of the Fermi surface.

Interacting Dirac fermions in graphene are described by the Hamiltonian ($\hbar = 1$)~\cite{GusyninSharapovCarbotte2007}
\begin{align}
H_{\mu} = &- \int_{\vec{r}} \Psi^{\dagger}(\vec{r}) \left( \mu + i v_F \sigma_0^s \otimes \vec{\Sigma} \cdot \vec{\nabla} \right) \Psi(\vec{r}) \nonumber \\
&+ \frac{1}{2} \int_{\vec{r}, \vec{r}'} \delta n_{\mu}(\vec{r}) \frac{e^2}{|\vec{r} - \vec{r}'|} \delta n_{\mu}(\vec{r}') \,,
\label{eq:GrandCanonicalHamiltonian}
\end{align}
with $\delta n_{\mu}(\vec{r}) = \Psi^{\dagger}(\vec{r}) \Psi(\vec{r}) - \tilde{n}_{\mu}(\vec{r})$.
%
%\begin{equation}
%\delta n_{\mu}(\vec{r}) = \Psi^{\dagger}(\vec{r}) \Psi(\vec{r}) - \tilde{n}_{\mu}(\vec{r}) \,.
%\label{eq:DefinitionDeltaN}
%\end{equation}
%
The Dirac electrons are described by eight-dimensional spinors $\Psi \equiv \begin{pmatrix} \Psi_{\uparrow} & \Psi_{\downarrow} \end{pmatrix}^{\intercal}$, with $\Psi_{\sigma} \equiv \begin{pmatrix} \psi_{A K_+} & \psi_{B K_+} & \psi_{B K_-} & \psi_{A K_-} \end{pmatrix}_{\sigma}^{\intercal}$.
%
%\begin{equation}
%\Psi_{\sigma} \equiv \begin{pmatrix} \psi_{A K_+} & \psi_{B K_+} & \psi_{B K_-} & \psi_{A K_-} \end{pmatrix}_{\sigma}^{\intercal} \,.
%\label{eq:BasisPsi}
%\end{equation}
%
The indices $\sigma = \uparrow, \downarrow$ denote the spin, $K_{\pm}$ the valley- and $A/B$ the sublattice degree of freedom. Furthermore, $\sigma_0^s$ is the two-dimensional unit matrix acting in spin space and $\Sigma_{1, 2} = \tau_3 \otimes \sigma_{1, 2}$ are four-dimensional matrices, with the Pauli matrices $\tau_3$ and $\sigma_{1, 2}$ acting in valley and sublattice space, respectively. The term $\tilde{n}_{\mu}(\vec{r})$ is a background charge density, which depends implicitly on the chemical potential. It represents the charge accumulated on a nearby metal gate and removes the zero wavenumber singularity of the bare Coulomb interaction.

The key insight of our method is that the chemical potential in Eq.~\eqref{eq:GrandCanonicalHamiltonian} couples to a fermion bilinear in exactly the same way as an additive infrared regulator in the fRG. Since the chemical potential is a continuous and differentiable variable it may formally be reinterpreted as a flow parameter.\cite{BergesEtal2003} This interpretation enables us to derive an exact flow equation for the chemical-potential-dependent effective action $\Gamma_{\mu}$ and to apply the by now well-established methods of the fRG. Since the essential steps to arrive at an exact flow equation are identical to the fRG, we can immediately transfer the general (finite temperature and density) fRG equations from Ref.~[\onlinecite{FrasdorfMosig2017}], taking care that the regularization prescription is substituted appropriately. In Fig.~\ref{GraphicalFlowEquations} we show a graphical representation of the flow equations for the one-particle irreducible vertex functions of the theory in its Fermi-Bose form. The bosonic field was introduced by a Hubbard-Stratonovich transformation of the Coulomb interaction in the density-density channel.~\cite{BauerKopietz2015, SchwieteFinkelstein2014-1, NegeleOrlandBook, KamenevBook, KopietzBook, FrasdorfMosig2017}
\begin{figure}
\begin{align*}
\partial_{\mu} \boldsymbol{\hat{\Sigma}}_{\mu} = i \partial \!\!\!/_{\mu}
\raisebox{-.8cm}{
\scalebox{.5}{
	\begin{fmfgraph}(40,25)
		\fmfleft{i}
		\fmfright{o}
		\fmf{fermion,tension=3}{G1,i}
		\fmfrpolyn{filled=15}{G}{3}
		\fmfpolyn{filled=15}{K}{3}
		\fmf{fermion,tension=1.5}{K3,G3}
		\fmf{wiggly,left=.5,tension=0}{G2,K2}
		\fmf{fermion,tension=3}{o,K1}
	\end{fmfgraph}
}} &\quad
\partial_{\mu} \boldsymbol{\Pi}_{\mu} = \frac{i}{2} \partial \!\!\!/_{\mu}
\scalebox{.5}{
	\begin{fmfgraph}(45,2.5)
		\fmfleft{i}
		\fmfright{o}
		\fmf{wiggly,tension=3}{i,G1}
		\fmfrpolyn{filled=15}{G}{3}
		\fmfpolyn{filled=15}{K}{3}
		\fmf{fermion,left=.3,tension=.5}{G2,K2}
		\fmf{fermion,left=.3,tension=.5}{K3,G3}
		\fmf{wiggly,tension=3}{K1,o}
	\end{fmfgraph}
}
\\[-.5cm]
\vspace{-2cm}
\partial_{\mu} \boldsymbol{\Gamma}_{\mu}^{(2,1)} &= i \partial \!\!\!/_{\mu}
\raisebox{-.5cm}{
\scalebox{.5}{
	\begin{fmfgraph}(25,25)
		\fmfleft{i1}
		\fmfright{o2,o3}
		\fmfpolyn{filled=15}{G}{3}
		\fmfpolyn{filled=15}{K}{3}
		\fmfpolyn{filled=15}{J}{3}
		\fmf{wiggly,tension=4}{G1,i1}
		\fmf{fermion,tension=4}{o2,K2}
		\fmf{fermion,tension=4}{J3,o3}
		\fmf{wiggly,tension=1}{K3,J2}
		\fmf{fermion,tension=1}{G3,J1}
		\fmf{fermion,tension=1}{K1,G2}
	\end{fmfgraph}
}}
\end{align*}
\caption{Graphical representation of the chemical potential flow equations for the self-energy $\boldsymbol{\hat{\Sigma}}_{\mu}$, the polarization function $\boldsymbol{\Pi}_{\mu}$, and the Fermi-Bose vertex $\boldsymbol{\Gamma}_{\mu}^{(2,1)}$ (shaded triangle). Contributions from higher order vertices $\boldsymbol{\Gamma}_{\mu}^{(m,n)}$, with $m>2, n>1$ are already neglected. Straight and wiggly lines represent the flowing fermionic and bosonic propagators, respectively. The derivative $\partial \!\!\!/_{\mu}$ on the right hand side only acts on the flowing fermionic propagators, substituting the latter by the so-called single scale propagator.~\cite{FrasdorfMosig2017}}
\label{GraphicalFlowEquations}
\end{figure}

In contrast to the standard fRG the main issue of concern in the chemical-potential flow theory is the initial condition of the flow. Since the chemical potential is different from an infrared regulator by its analytical structures, the effective action at some - arbitrarily chosen - initial chemical potential $\mu_0$ is nontrivial and, in particular, does not coincide with the bare action. It has to be calculated separately, using an appropriate nonperturbative method such as the fRG or Schwinger-Dyson equations.~\cite{PopoviciSmekal2013} Here, we have chosen to use the Keldysh-fRG framework~\cite{GezziPruschkeMeden2006, JakobsMedenSchoeller2007, GasenzerPawlowski2008, BergesHoffmeister2009, KarraschEtal2010, KlossKopietz2011, KennesEtal2012, BergesMesterhazy2012} we implemented in Ref.~[\onlinecite{FrasdorfMosig2017}]. In this work, we calculated the Fermi velocity and the static dielectric function as functions of momentum and temperature at zero carrier density. The truncation scheme in Ref.~[\onlinecite{FrasdorfMosig2017}] neglects any dynamical effects, such as plasmons and the quasiparticle wavefunction renormalization, the three-vertex renormalization and higher-order vertices entirely. We use these results as the starting point of the chemical-potential flow. 

To be consistent with the fRG calculation, we employ the same level of truncation and the same approximations for the chemical-potential based flow. That means, in particular, we limit ourselves to the flow of the quasiparticle pole (temperature arguments are suppressed throughout)
\begin{equation}
\xi_{\mu}(k) = v_{\mu}(k) k \,,
\label{eq:DefinitionXi}
\end{equation}
%
%where the renormalized and $\mu$-dependent Fermi velocity $v_{\mu}(k)$ has been defined as $v_{\mu}(k) = v_F + \Sigma_{v, \mu}(k)$,
%
%\begin{equation}
%v_{\mu}(k) = v_F + \Sigma_{v, \mu}(k) \,.
%\label{eq:DefinitionRenormalizedFermiVelocity}
%\end{equation}
%
and the flow of the static dielectric function
\begin{equation}
\epsilon_{\mu}(\vec{q}) \equiv \epsilon_0 \left( 1 + V(\vec{q}) \Pi_{\mu}^{R/A}(\omega = 0, \vec{q}) \right) \,.
\label{eq:DefinitionDielectricFunction}
\end{equation}
Here, the renormalized and $\mu$-dependent Fermi velocity $v_{\mu}(k)$ has been defined as $v_{\mu}(k) = v_F + \Sigma_{v, \mu}(k)$, and $V(\vec{q}) = 2 \pi e^2 /q$ is the Fourier transform of the bare Coulomb interaction.
Assuming the absence of spontaneous chiral symmetry breaking, we obtain two coupled flow equations from the general vertex flow equations shown in Fig.~\ref{GraphicalFlowEquations}, one for the Fermi velocity $v_{\mu}(k)$ and one for the static dielectric function $\epsilon_{\mu}(q)$, see the appendix for details. For convenience we introduce the function $\chi_{\mu}(q) \equiv \epsilon_{\mu}(q) q$, which is - up to constants - the inverse of the renormalized Coulomb interaction, and state the two flow equations in terms of $\xi_{\mu}(k)$ and $\chi_{\mu}(q)$, ($k_B = 1$)
\begin{widetext}
\begin{align}
\partial_{\mu} \xi_{\mu}(k) = \frac{e^2}{4 \pi} \int_0^{\infty} dq \int_0^{\pi} d\varphi \, \frac{1}{2T} \left( \textrm{cosh}^{-2} \left( \frac{\xi_{\mu}(q) + \mu}{2 T} \right) - \textrm{cosh}^{-2} \left( \frac{\xi_{\mu}(q) - \mu}{2 T} \right) \right) \frac{q \textrm{cos} \varphi}{\chi_{\mu} \left(\sqrt{q^2 + k^2 - 2 q k \, \textrm{cos} \varphi} \right)} \,,
\label{eq:FlowXiFiniteT}
\end{align}
\begin{align}
\partial_{\mu} \chi_{\mu}(q) = - \frac{e^2 q^2}{4 \pi T} \int_0^{\infty} d\rho \int_0^{\pi} d\phi \, \Bigg[ &\sum_{\nu = \pm} \nu \left( \textrm{cosh}^{-2} \left( \frac{\xi_{\mu}(\mathcal{Q}_-) - \nu \mu}{2 T} \right) + \textrm{cosh}^{-2} \left( \frac{\xi_{\mu}(\mathcal{Q}_+) - \nu \mu}{2 T} \right) 
%+ \textrm{tanh} \left( \frac{\xi_{\mu}(\mathcal{Q}_-) + \mu}{2 T} \right) + \textrm{tanh} \left( \frac{\xi_{\mu}(\mathcal{Q}_+) + \mu}{2 T} \right)
\right) \frac{\textrm{sin}^2 \phi}{\xi_{\mu}(\mathcal{Q}_-) + \xi_{\mu}(\mathcal{Q}_+)} \nonumber \\
+ &\sum_{\nu = \pm} \nu \left( \textrm{cosh}^{-2} \left( \frac{\xi_{\mu}(\mathcal{Q}_-) - \nu \mu}{2 T} \right) - \textrm{cosh}^{-2} \left( \frac{\xi_{\mu}(\mathcal{Q}_+) - \nu \mu}{2 T} \right) 
%+ \textrm{tanh} \left( \frac{\xi_{\mu}(\mathcal{Q}_-) + \mu}{2 T} \right) - \textrm{tanh} \left( \frac{\xi_{\mu}(\mathcal{Q}_+) + \mu}{2 T} \right) 
\right) \frac{\textrm{sinh}^2 \rho}{\xi_{\mu}(\mathcal{Q}_-) - \xi_{\mu}(\mathcal{Q}_+)} \Bigg] \,.
\label{eq:FlowChiFiniteT}
\end{align}
\end{widetext}
Here, $\mathcal{Q}_{\pm}$ is a short hand notation for the function $\mathcal{Q}_{\pm}(\rho, \phi, q) = \tfrac{1}{2} q \, (\textrm{cosh} \, \rho \pm \textrm{cos} \, \phi)$, where $\rho$ and $\phi$ are elliptic coordinates, and the summation over $\nu = \pm$ covers the valence and conduction band. 
In the limit of vanishing temperature the inverse hyperbolic cosine is proportional to a delta function, centered at the interacting Fermi surface $\xi_{\mu}(k_F) \pm \mu = 0$.~\cite{Note1}
%~\footnote{Depending on the sign of $\mu$, the interacting Fermi surface, being a circle of radius $k_F$, is located either within the conduction or valence band.} 
For finite, not too large temperatures the delta-function singularity is smeared out, but remains strongly peaked at the Fermi surface, whereas those modes for which $\xi_{\mu}(q) \pm \mu \gg 2T$ are exponentially suppressed. Hence, the momentum integrals of the two flow equations are finite, both in the ultraviolet and infrared regime. We note that the above flow equations are fully symmetric with respect to the sign of $\mu$. This fact is a consequence of the chiral symmetry of the model~\eqref{eq:GrandCanonicalHamiltonian} and the assumed absence of symmetry breaking. Hence, without loss of generality we may consider positive $\mu$, \textit{i.e.} $n$-doping.

\begin{figure}
	\includegraphics[width=.49\textwidth]{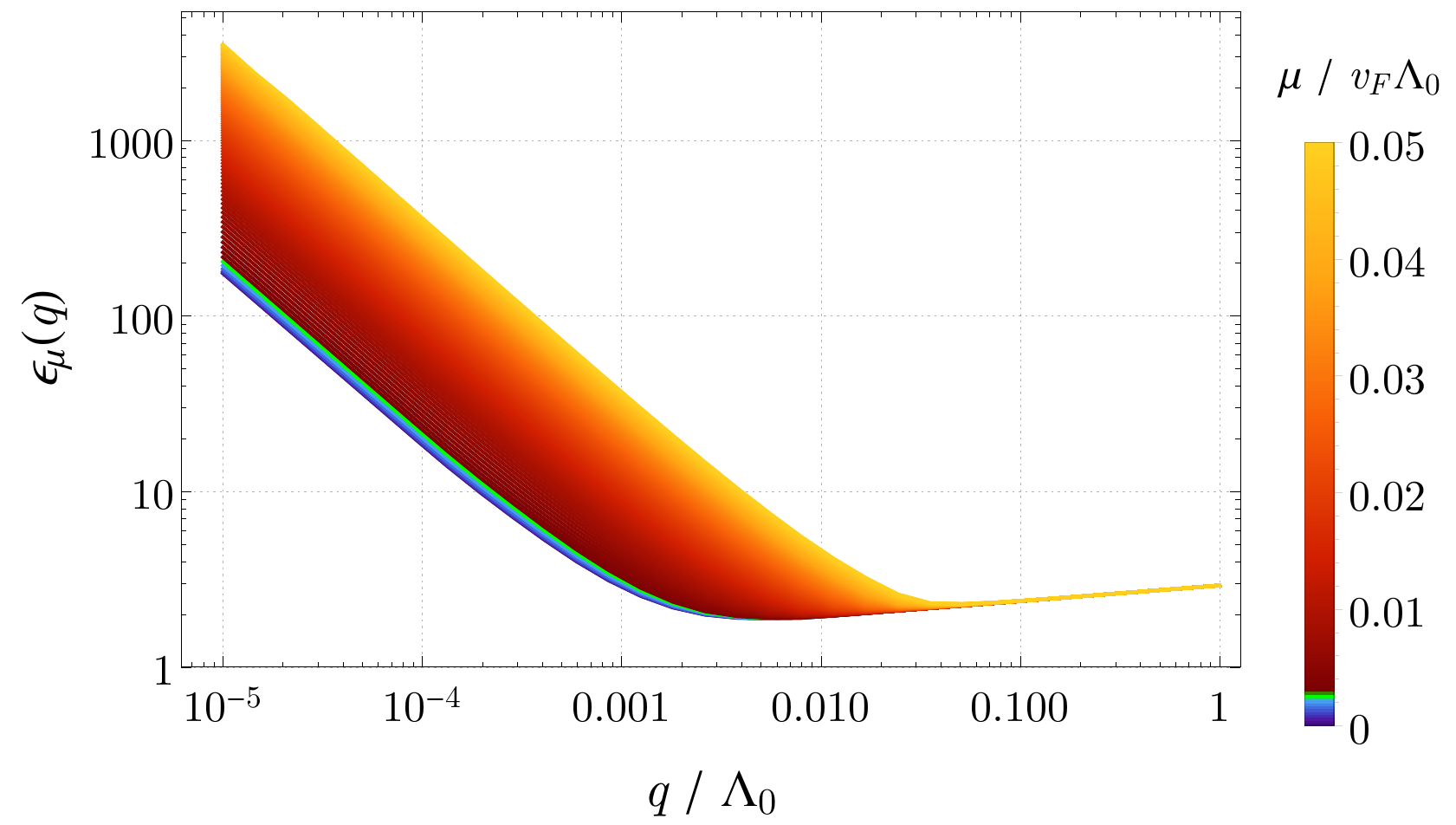}
	\caption{(Color online) Dielectric function $\epsilon_{\mu}(q)$ as a function of momentum and chemical potential at the reduced temperature $T/v_F \Lambda_0 = 2.5 \times 10^{-3}$. The colors blue to green and red to orange separate the two regimes $\mu \leq T$ and $\mu > T$, respectively. At the charge neutrality point the long range tail of the bare Coulomb interaction is cut off, due to thermal screening.}
	\label{fig:Epsilon_Full_T0-0025_2D}
\end{figure}

The flow equations \eqref{eq:FlowXiFiniteT} and \eqref{eq:FlowChiFiniteT} have been solved numerically for different temperatures with the dimensionless coupling constant $\alpha = e^2/v_F = 2.2$ appropriate for freestanding graphene.\cite{KotovEtal2012, BauerKopietz2015} The flow has been initialized at the charge neutrality point $\mu_0 = 0$, with nonperturbative initial conditions 
\begin{equation}
\xi_{\mu_0 = 0}^{\textrm{fRG}}(k) = v_{\mu_0 = 0}^{\textrm{fRG}}(k) k \,, \quad \chi_{\mu_0 = 0}^{\textrm{fRG}}(q) = \epsilon_{\mu_0 = 0}^{\textrm{fRG}}(q) q \,,
\label{eq:FRGInitialConditions}
\end{equation}
which we have obtained by an fRG calculation.~\cite{FrasdorfMosig2017} 
%The raw data for the Fermi velocity have been smoothened in the momentum regime $q \lesssim T/v_F$ before insertion into the chemical-potential flow. In this momentum regime the data showed small fluctuations around a mean value, which we attributed to numerical uncertainties rather than a real physical effect. For the temperature presented here, the absolute deviation from the mean was of the order $10^{-2}$. 
%
The numerical results for the dielectric function are shown in Fig.~\ref{fig:Epsilon_Full_T0-0025_2D} for the reduced temperature $T/v_F \Lambda_0 = 2.5 \times 10^{-3}$, where $\Lambda_0$ is the upper band cutoff of the low-energy Hamiltonian~\eqref{eq:GrandCanonicalHamiltonian}.

\begin{figure}
\includegraphics[width=0.49\textwidth]{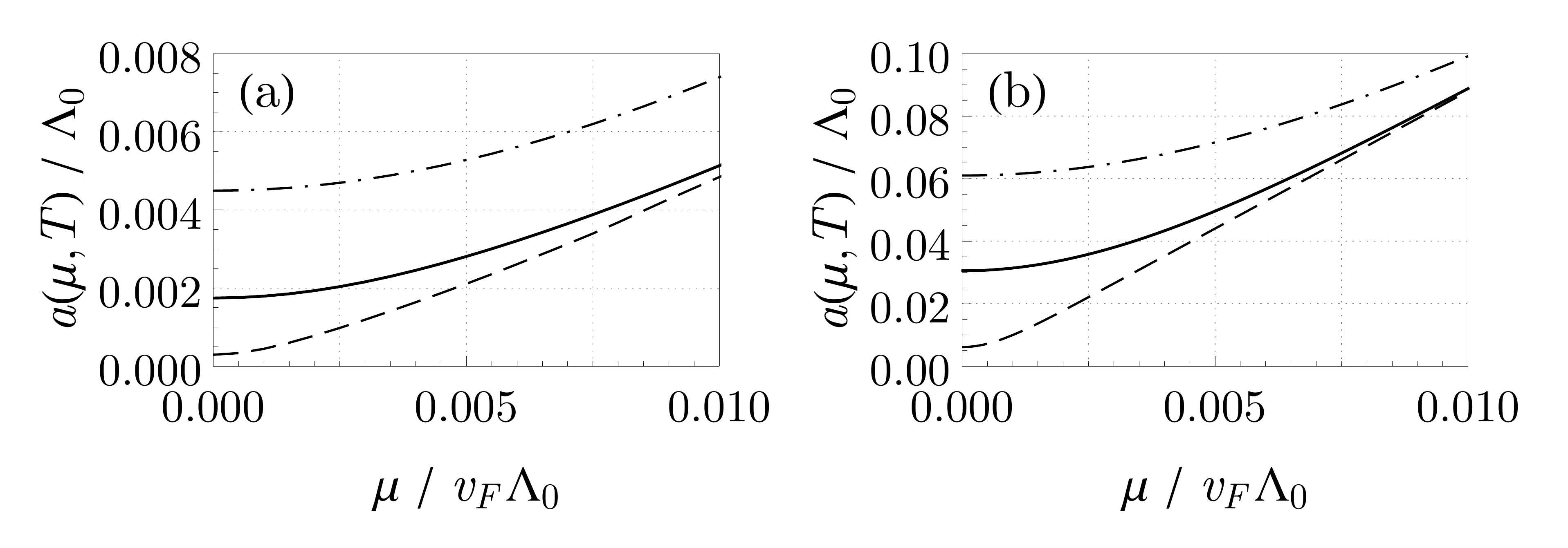}
%	\subfigure[]{\includegraphics[width=0.235\textwidth]{a(muT)Full.pdf}} \hfill
%	\subfigure[]{\includegraphics[width=0.235\textwidth]{a(muT)OneLoop.pdf}}
	\caption{Quantitative comparison between the (a) fully self-consistent solution and (b) one loop approximation of the coefficient function $a(\mu, T)$ for the reduced temperatures $T/v_F \Lambda_0 = 5 \times 10^{-4}, 2.5 \times 10^{-3}, 5 \times 10^{-3}$ (bottom to top data sets). In the small momentum regime, $q \ll T/v_F, \mu/v_F$, the static dielectric function shows a $1/q$ divergence according to $\epsilon_{\mu}(q) = 1 + a(\mu, T) \Lambda_0/q$. Observe that the self-consistent solution is about an order of magnitude smaller than the one-loop prediction.}
	\label{fig:a_Full_OneLoop}
\end{figure}

At the charge neutrality point the dielectric function shows a distinctively different behaviour in the two momentum regimes $q > T/v_F$ and $q < T/v_F$. While the dielectric function is only weakly dependent on the momentum in the regime $q \gg T/v_F$, a strong $1/q$ divergence can be observed for  $q \ll T/v_F$. As explained in Ref.~[\onlinecite{FrasdorfMosig2017}], this divergence could be attributed to thermally induced charge carriers. In the presence of a finite chemical potential, that is excess charge carriers, the large momentum components of the dielectric function remain unaffected, whereas the initial $1/q$ divergence found in the low momentum regime becomes strongly enhanced, leading to an increasingly short ranged renormalized Coulomb interaction. This picture is consistent with the results obtained in one-loop perturbation theory. For comparison, in the regime $q \ll T/v_F$ perturbation theory predicts a polarization function, that - in the static limit - is independent of momentum and a function of temperature and chemical potential only,~\cite{SchuettGornyiMirlin2011}
\begin{equation}
\epsilon_{1-\textrm{loop}}(q) = 1 + a(\mu, T) \frac{\Lambda_0}{q} \,,
\label{eq:EpsilonOneLoop}
\end{equation}
with
\begin{equation}
a(\mu, T) = 8 \alpha \frac{T}{v_F \Lambda_0} \textrm{ln} \left( 2 \textrm{cosh} \frac{\mu}{2 T} \right) \,.
\label{eq:a(mu,T)-1Loop}
\end{equation}
Here, the coefficient function $a(\mu, T)$ is directly proportional to the static limit of the polarization function. At the charge neutrality point $a(\mu, T)$ scales linearly with temperature, whereas for $\mu \gg T$ it becomes independent of temperature, scaling linearly with the chemical potential. The former feature has been shown to remain valid in a nonperturbative fRG calculation,~\cite{FrasdorfMosig2017} showing a strong renormalization of the slope. To verify whether the latter feature remains valid beyond perturbation theory, we solved the self-consistency Eqs.~\eqref{eq:FlowXiFiniteT} and \eqref{eq:FlowChiFiniteT} for the two additional temperatures $T/v_F \Lambda_0 = 5 \times 10^{-4}, 5 \times 10^{-3}$ and extracted the coefficient functions $a(\mu, T)$, see Fig.~\ref{fig:a_Full_OneLoop}.
For large chemical potentials we observed a transition into a linear regime, which is consistent with the result obtained by perturbation theory. However, the precise slope could not be determined sufficiently accurate due to convergence issues of the numerical integration: At very small momenta and increasingly large chemical potentials the integrand of Eq.~\eqref{eq:FlowChiFiniteT} becomes very strongly peaked, such that limited machine-precision becomes problematic. 
%Decreasing the temperature worsened this problem as it reduced the chemical potentials that could be accessed. 
Nevertheless, our results strongly indicate that the scaling behaviour predicted by perturbation theory is indeed correct, albeit with a strongly renormalized slope. A precise estimation of the slope would require a recalculation of the temperature dependence of the renormalized Fermi velocity and dielectric function at the charge neutrality point with a better resolution and accuracy than what was achieved previously in Ref.~[\onlinecite{FrasdorfMosig2017}].

\begin{figure}
\includegraphics[width=0.49\textwidth]{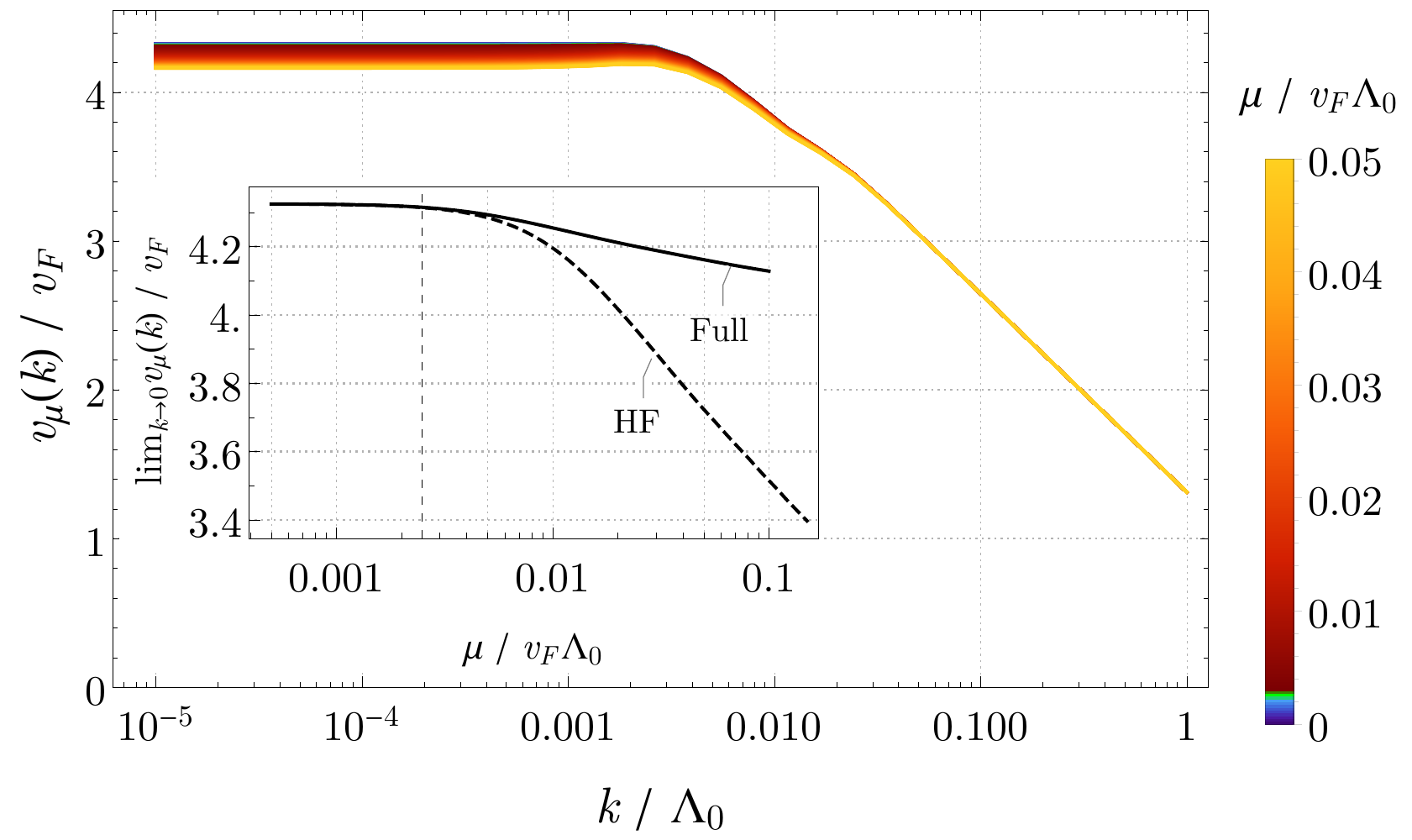}
%	\subfigure[]{\includegraphics[width=0.49\textwidth]{v_Full_T0-0025_Inset.pdf}} \hfill
	%\subfigure[]{\includegraphics[width=0.23\textwidth]{HartreeFockTest.pdf}}
	%\subfigure[]{\includegraphics[width=0.47\textwidth]{v(0)_Full_T0-0025.pdf}}
	\caption{(Color online) Renormalized Fermi velocity $v_{\mu}(k)$ as a function of momentum and chemical potential at temperature $T/v_F \Lambda_0 = 2.5 \times 10^{-3}$ in the fully self-consistent calculation. The renormalized Fermi velocity is finite at $k = 0$ at the initial chemical potential $\mu_0 = 0$ due to temperature induced screening of the renormalized Coulomb interaction. Increasing the chemical potential away from the charge neutrality point shows a weak suppression of the Fermi velocity for small momenta.
The inset shows a comparison between the $k \rightarrow 0$ limits of the Fermi velocity in the self-consistent treatment and in Hartree-Fock approximation as functions of the chemical potential. For $\mu > 2T$ both calculations show a logarithmic suppression of $v_{\mu}(k \rightarrow 0)$. This suppression is significantly weakened in the full computation when compared to the result of the Hartree-Fock approximation.
}
	\label{fig:v-k-mu}
\end{figure}

The numerical results for the chemical potential dependence of the renormalized Fermi velocity are shown in Fig.~\ref{fig:v-k-mu}. At the initial chemical potential $\mu_0 = 0$ the infrared divergence of the renormalized Fermi velocity is regularized due to the temperature-induced screening of the renormalized Coulomb interaction. Upon increasing the chemical potential the solution shows a further, but only very weak suppression of the Fermi velocity at low momenta in accord with the assumption of Bauer \textit{et al.}\cite{BauerKopietz2015}
Our full calculation allows us to understand this behaviour by considering the combined effect of strong screening and the formation of a nontrivial Fermi surface. By increasing the chemical potential the additional charge carriers fill up the renormalized spectrum and introduce a circularly shaped Fermi surface, which is driven further and further away from the nodal point, while the renormalized Coulomb interaction becomes increasingly short ranged. As a result, the screened Coulomb interaction only operates near the Fermi surface and, loosely speaking, does not reach far enough into the spectrum to have a significant impact on the small momentum regime of the renormalized Fermi velocity. Neglecting the charge carrier induced screening would cause a much stronger suppression of the Fermi velocity, since then the Coulomb interaction could reach down to the nodal point. In order to validate this picture we also performed a Hartree-Fock like calculation of the velocity, see inset of Fig.~\ref{fig:v-k-mu}, where only Eq.~\eqref{eq:FlowXiFiniteT} has been solved self-consistently for $\xi_{\mu}(k)$. The $\mu$ flow of the dielectric function therein was neglected and the function $\chi_{\mu}(q) = \epsilon_{\mu}(q) q$ was kept at its initial value $\chi_{\mu}(q) = \chi_{\mu_0 = 0}(q)$, where only temperature induced screening is present. The Hartee-Fock solution shows the same features as the fully self-consistent solution. However, the low momentum regime of the Hartree-Fock Fermi velocity is much stronger suppressed, supporting the above reasoning.

The idea to use the chemical potential as a flow parameter in a functional renormalization group calculation was first put forward by Berges \textit{et al.}~\cite{BergesEtal2003} in the context of a particle-physics problem. We here have shown that the fRG with the chemical potential as the flow parameter is an efficient tool for applications in condensed matter physics. Although a separate nonperturbative calculation is required to establish the initial condition for the flow, this initial ``investment'' pays off, because for a chemical-potential based flow each point in the solution of the flow equation is of physical relevance, in contrast to more standard fRG approaches, where only the end point of the flow matters. The alternative would be to run a conventional fRG calculation for each value of the chemical potential separately, which should yield comparable results, but involves much greater effort.

We have applied the technique for the calculation of the carrier density dependence of the Fermi velocity and the static dielectric function in graphene, using a conventional fRG calculation at zero chemical potential as initial condition. Graphene is a very suitable context for an application of the chemical-potential flow technique. In this material physical quantities in principle have a strong dependence of the carrier density, providing a need for such a calculation, while the high-symmetry point of zero carrier density brings significant simplifications, allowing for an efficient ``conventional'' nonperturbative calculation at that point. Our numerical results fully support the earlier work of Bauer \textit{et al.},~\cite{BauerKopietz2015} which took the momentum-dependent Fermi velocity to be independent of the chemical potential. In particular, their theoretical fit to the experimental data of Elias \textit{et.al.}~\cite{EliasEtal2011} can be justified by the results of this paper. The dielectric function is, however, strongly dependent on the chemical potential, reflecting the strong carrier dependence of the screening length in graphene. It would be interesting to apply our method to the three dimensional analog of graphene, the so-called Weyl semi-metals. Such materials feature a conical spectrum as well and a chemical-potential based flow could be implemented equally efficiently.

We thank Piet Brouwer and Bj\"orn Sbierski for support in the preparation of the manuscript. This work is supported by the German Research Foundation (DFG) in the framework of the Priority Program 1459 ``Graphene''.

\section{Appendix}
\label{sec:Appendix}

In this appendix we explain some details about the general chemical potential flow theory, which is the basis for the flow equations \eqref{eq:FlowXiFiniteT} and \eqref{eq:FlowChiFiniteT}. Since the theory relies on a reinterpretation of the chemical potential as a flow parameter, the main results can be transfered directly from Ref.~[\onlinecite{FrasdorfMosig2017}]. %As a consequence we present the main ingredients of the theory as brief as possible.

The starting point for the derivation of an exact flow equation is the $\mu$-dependent partition function $Z_{\mu}[\boldsymbol{\eta}, \boldsymbol{J}]$. It is defined as the functional Fourier transform of the exponentiated bare action $S_{\mu}[\boldsymbol{\psi}, \boldsymbol{\phi}]$,~\cite{BergesMesterhazy2012, KamenevBook, NegeleOrlandBook, KopietzBook}
%
%\begin{align}
%Z_{\mu}[\boldsymbol{\eta}, \boldsymbol{J}] &= \left\langle e^{i \boldsymbol{\eta^{\dagger}} \tau_1 \boldsymbol{\Psi} + i \boldsymbol{\Psi^{\dagger}} \tau_1 \boldsymbol{\eta} + i \boldsymbol{\phi}^{\intercal} \tau_1 \boldsymbol{J}} \right\rangle \,, 
%\label{eq:PartitionFunctionA} \\
%\langle \cdots \rangle &= \int \mathcal{D}\psi \mathcal{D}\psi^{\dagger} \mathcal{D}\phi \cdots e^{iS_{\mu}[\boldsymbol{\psi}, \boldsymbol{\phi}]} \,,
%\label{eq:PartitionFunctionB}
%\end{align}
%
\begin{align}
Z_{\mu}[\boldsymbol{\eta}, \boldsymbol{J}] = \int\! \mathcal{D}\psi \mathcal{D}\psi^{\dagger} \mathcal{D}\phi \, e^{iS_{\mu}[\boldsymbol{\psi}, \boldsymbol{\phi}] + i \boldsymbol{\eta^{\dagger}} \tau_1 \boldsymbol{\Psi} + i \boldsymbol{\Psi^{\dagger}} \tau_1 \boldsymbol{\eta} + i \boldsymbol{\phi}^{\intercal} \tau_1 \boldsymbol{J}}.
\label{eq:PartitionFunction}
\end{align}
The bare action is a functional of fermionic and bosonic fields, which can be derived from the purely fermionic Hamiltonian~\eqref{eq:GrandCanonicalHamiltonian} by a standard procedure.~\cite{SchwieteFinkelstein2014-1, NegeleOrlandBook, KamenevBook, FrasdorfMosig2017} The bosonic field is introduced by a Hubbard-Stratonovich transformation of the Coulomb interaction term in the density-density channel. The index $\mu$ indicates that both the partition function and the bare action depend on the chemical potential. The chemical potential dependence of the bare action enters explicitly via the quadratic $\mu$-term in the Hamiltonian and implicitly via the background density $\tilde{n}_{\mu}$. In contrast to the conventional fRG there is no additional infrared regulator.~\cite{BergesEtal2003} Furthermore, we work in the real-time Keldysh formalism,~\cite{GezziPruschkeMeden2006, JakobsMedenSchoeller2007, GasenzerPawlowski2008, BergesHoffmeister2009, KarraschEtal2010, KlossKopietz2011, KennesEtal2012, BergesMesterhazy2012} which involves a doubling of degrees of freedom, with classical $(c)$ and quantum $(q)$ component for each field~\cite{SchwieteFinkelstein2014-1, BergesMesterhazy2012, FrasdorfMosig2017, KamenevBook}
\begin{equation}
\boldsymbol{\Psi} \equiv \begin{pmatrix} \Psi_c & \Psi_q \end{pmatrix}^{\intercal} \,, \quad \boldsymbol{\Psi^{\dagger}} = (\boldsymbol{\Psi})^{\dagger} \,, \quad
\boldsymbol{\phi} \equiv \begin{pmatrix} \phi_c & \phi_q \end{pmatrix}^{\intercal} \,.
\label{eq:DefinitionClassicalQuantumFields}
\end{equation}
Lastly, $\boldsymbol{\eta}$ and $\boldsymbol{J}$ are fermionic and bosonic source fields, respectively. In Eq.%~\eqref{eq:PartitionFunctionA}
~\eqref{eq:PartitionFunction}
we employed a condensed vector notation for the source terms, containing integration and summation of continuous and discrete field degrees of freedom implicitly, \textit{e.g.}
\begin{equation}
\boldsymbol{\eta^{\dagger}} \tau_1 \boldsymbol{\Psi} \equiv \int_x \boldsymbol{\eta^{\dagger}}(x) \tau_1 \boldsymbol{\Psi}(x) \,,
\label{eq:CondensedNotation}
\end{equation}
where $\tau_1$ is a Pauli matrix acting in Keldysh space.

The effective action may now be introduced as the modified Legendre transform of the connected functional $W_{\mu}[\boldsymbol{\eta}, \boldsymbol{J}] = -i \textrm{ln} Z_{\mu}[\boldsymbol{\eta}, \boldsymbol{J}]$,\cite{Wetterich2001, BergesTetradisWetterich2002, BergesEtal2003, MetznerHonercampEtal2012, KopietzBook}
\begin{align}
\Gamma_{\mu}[\boldsymbol{\psi}, \boldsymbol{\phi}] = \, &W_{\mu}[\boldsymbol{\eta}_{\mu}, \boldsymbol{J}_{\mu}] - \boldsymbol{\eta^{\dagger}}_{\mu} \tau_1 \boldsymbol{\Psi} - \boldsymbol{\Psi^{\dagger}} \tau_1 \boldsymbol{\eta}_{\mu} - \boldsymbol{\phi}^{\intercal} \tau_1 \boldsymbol{J}_{\mu} \nonumber \\
&- \boldsymbol{\Psi}^{\dagger} \boldsymbol{\hat{R}}_{\mu} \boldsymbol{\Psi} \,,
\label{eq:LegendreTransform}
\end{align}
with
\begin{align}
\boldsymbol{\hat{R}}_{\mu}(x, y) &= \begin{pmatrix} 0 & \mu \, \delta(x - y) \hat{\mathbb{1}} \\ \mu \, \delta(x - y) \hat{\mathbb{1}} & 0 \end{pmatrix} \,.
\label{eq:KeldyshStructureFermionicRegulators}
\end{align}
The term $\boldsymbol{\Psi}^{\dagger} \boldsymbol{\hat{R}}_{\mu} \boldsymbol{\Psi}$ is the explicit chemical-potential term one obtains in the bare action $S_{\mu}[\boldsymbol{\psi}, \boldsymbol{\phi}]$. Its resemblance with an additive infrared regulator in the conventional fRG is the foundation of the chemical-potential flow theory.~\cite{BergesEtal2003} According to the usual definition of the effective flowing action the ``chemical-potential regulator term'' has been subtracted on the right hand side. Consequently the effective action $\Gamma_{\mu}$ involves flowing vertex functions only, and the explicit chemical-potential term - in comparison to the bare action - is absent. Note that some authors prefer to include a finite chemical potential in the fermionic distribution function, rather than in the spectral part of the inverse propagators as we do here.~\cite{KamenevBook} Such an alternative choice would affect the structure of the regulator~\eqref{eq:KeldyshStructureFermionicRegulators} and the vertex expansion of the effective action, but it cannot lead to any observable consequences, since these two choices are connected by a (time dependent) gauge transformation. 

The exact chemical potential flow equation follows immediately upon taking the $\mu$-derivative of Eq.~\eqref{eq:LegendreTransform}, keeping the fields $\boldsymbol{\psi}$ and  $\boldsymbol{\phi}$ fixed
\begin{align}
\partial_{\mu} \Gamma_{\mu}[\boldsymbol{\psi}, \boldsymbol{\phi}] = \frac{i}{2} \partial \!\!\!/_{\mu} \textrm{STr} \, \textrm{ln} \left( \boldsymbol{\hat{\Gamma}}_{\mu}^{(2)}[\boldsymbol{\psi}, \boldsymbol{\phi}] + \boldsymbol{\hat{\mathcal{R}}}_{\mu} \right) + 2 \phi_q \partial_{\mu} \tilde{n}_{\mu} \,,
\label{eq:ExactFlowEquation}
\end{align}
where $\boldsymbol{\hat{\Gamma}}_{\mu}^{(2)}$ is a Hesse matrix of second derivatives, and
\begin{equation}
\boldsymbol{\hat{\mathcal{R}}}_{\mu} = \textrm{diag} \left( - \boldsymbol{\hat{R}}_{\mu}, \boldsymbol{\hat{R}}_{\mu}^{\intercal}, 0 \right) \,.
\label{eq:RegulatorMatrix}
\end{equation}
The ``single-scale derivative'' $\partial \!\!\!/_{\mu}$ in Eq.~\eqref{eq:ExactFlowEquation} only acts on the regulator $\boldsymbol{\hat{\mathcal{R}}}_{\mu}$. The above flow equation is the Keldysh analog of the original, imaginary-time flow equation proposed by Berges \textit{et al.}~\cite{BergesEtal2003} The flow equations for the one-particle irreducible vertex functions are obtained by expanding the effective action in powers of fields, which needs to be inserted into the above equation, and comparing coefficients. Since we are only interested in thermal equilibrium, where the fluctuation-dissipation theorem holds,~\cite{KamenevBook} we only need to consider the resulting flow equations for the retarded components of the self-energy and polarization function. In a condensed notation, where numerical arguments denote space and time coordinates, $1 \equiv (\vec{r}_1, t_1)$, and latin indices encompass the discrete fermionic degrees of freedom, sublattice, valley and spin, these flow equations read
\begin{widetext}

\begin{align}
\partial_{\mu} \Sigma_{\mu, ij}^R(1, 2) = i \partial \!\!\!/_{\mu} \sum_{k, l} \int\nolimits'\ \Big( &\Gamma_{\mu, ik}^{qcc}(1, 1'; 4') G_{\mu, kl}^K(1', 2') \Gamma_{\mu, lj}^{ccq}(2', 2; 3') D_{\mu}^A(3', 4') \nonumber \\ 
+ &\Gamma_{\mu, ik}^{qcc}(1, 1'; 4') G_{\mu, kl}^R(1', 2') \Gamma_{\mu, lj}^{qcc}(2', 2; 3') D_{\mu}^K(3', 4') \Big) \,,
\label{eq:FlowEqRetardedSelfenergy}
\end{align}
\begin{align}
\partial_{\mu} \Pi_{\mu}^R(1, 2) = \frac{i}{2} \partial \!\!\!/_{\mu} \sum_{k, l, m, n} \int\nolimits'\ \Big( &G_{\mu, kl}^K(1', 2') \Gamma_{\mu, lm}^{ccq}(2', 3'; 1) G_{\mu, mn}^R(3', 4') \Gamma_{\mu, nk}^{qcc}(4', 1'; 2) \nonumber \\
+ &G_{\mu, kl}^A(1', 2') \Gamma_{\mu, lm}^{ccq}(2', 3'; 1) G_{\mu, mn}^K(3', 4') \Gamma_{\mu, nk}^{cqc}(4', 1'; 2) \Big) \,.
\label{eq:FlowEqRetardedPolarization}
\end{align}

\end{widetext}
The functions $\Gamma_{\mu, ij}^{\alpha \beta \gamma}(1, 2; 3)$, with $\alpha, \beta, \gamma = c, q$ are the Fermi-Bose three-vertices of the theory in the real-time Keldysh formulation. The primed integration sign indicates that all primed arguments have to be integrated. A transformation to Fourier-space is beneficial, due to energy and momentum conservation. The flowing frequency-momentum space propagators that enter the transformed flow equations read
\begin{subequations}
\begin{align}
\hat{G}_{\mu}^{R/A}(\vec{k}, \varepsilon) &= \frac{1}{\sigma_0^s \otimes \left(\varepsilon + \mu - \vec{\Sigma} \cdot \vec{k} - \hat{\Sigma}_{\mu}^{R/A}(\vec{k}, \varepsilon) \right)} \,, \\
\hat{G}_{\mu}^{K}(\vec{k}, \varepsilon) &= \textrm{tanh} \frac{\varepsilon}{2T} \left( \hat{G}_{\mu}^R(\vec{k}, \varepsilon) - \hat{G}_{\mu}^A(\vec{k}, \varepsilon) \right) \,,
\label{eq:FermionPropagators}
\end{align}
\end{subequations}
\begin{subequations}
\begin{align}
D_{\mu}^{R/A}(\vec{q}, \omega) &= \frac{1}{2} \frac{1}{V^{-1}(\vec{q}) + \Pi_{\mu}^{R/A}(\vec{q}, \omega)} \,, \\
D_{\mu}^K(\vec{q}, \omega) &= \textrm{coth} \frac{\omega}{2T} \left( D_{\mu}^R(\vec{q}, \omega) - D_{\mu}^A(\vec{q}, \omega) \right) \,.
\label{eq:BosonPropagators}
\end{align}
\end{subequations}
The single-scale derivative in Eqs.~\eqref{eq:FlowEqRetardedSelfenergy} and \eqref{eq:FlowEqRetardedPolarization} only acts on the flowing fermionic propagators, substituting the latter by a single-scale propagator
\begin{equation}
\partial \!\!\!/_{\mu} \hat{G}_{\mu}^{R/A}(\vec{k}, \varepsilon) = - \left( \hat{G}_{\mu}^{R/A}(\vec{k}, \varepsilon) \right)^2 \,.
\label{eq:SingleScalePropagator}
\end{equation}
Here, the $\mu$-dependence of the flowing self-energy is held constant upon taking the single-scale derivative $\partial \!\!\!/_{\mu}$. By using the approximations mentioned in the main text - that is setting all the three-vertices to unity and neglecting any dynamical effects - we arrive after a straightforward calculation at the flow equations \eqref{eq:FlowXiFiniteT} and \eqref{eq:FlowChiFiniteT}.

%%%%%%%%%%%%%%%%%%%%%%%%%%%%%%%%%%%%%%%%%%%%%%%%%%%%
% Bibliography
%%%%%%%%%%%%%%%%%%%%%%%%%%%%%%%%%%%%%%%%%%%%%%%%%%%%

%\bibliography{frg_graphene}

%merlin.mbs apsrev4-1.bst 2010-07-25 4.21a (PWD, AO, DPC) hacked
%Control: key (0)
%Control: author (8) initials jnrlst
%Control: editor formatted (1) identically to author
%Control: production of article title (-1) disabled
%Control: page (0) single
%Control: year (1) truncated
%Control: production of eprint (0) enabled
%

\end{fmffile}

\end{document}